# Constructing a Portfolio Optimization Benchmark Framework for Evaluating Large Language Models

Hanyong Cho
Graduate School of Management
of Technology, Korea University
whgksdyd1@korea.ac.kr

Jang Ho Kim
Graduate School of Management
of Technology, Korea University
janghokim@korea.ac.kr

## ABSTRACT

This study introduces a benchmark framework for evaluating the financial decision-making capabilities of large language models (LLMs) through portfolio optimization problems with mathematically explicit solutions. Unlike existing financial benchmarks that emphasize language-processing tasks, the proposed framework directly tests optimization-based reasoning in investment contexts. A large set of multiple-choice questions is generated by varying objectives, candidate assets, and investment constraints, with each problem designed to include a unique correct solution and systematically constructed alternatives. Experimental results comparing GPT-4, Gemini 1.5 Pro, and Llama 3.1-70B reveal distinct performance patterns: GPT achieves the highest accuracy in risk-based objectives and remains stable under constraints, Gemini performs well in return-based tasks but struggles under other conditions, and Llama records the lowest overall performance. These findings highlight both the potential and current limitations of LLMs in applying quantitative reasoning to finance, while providing a scalable foundation for developing LLM-based services in portfolio management.

## 1. Introduction

The rapid development of artificial intelligence has positioned large language models (LLMs) as powerful tools for supporting decision-making across a wide range of industries. As noted by Zhao et al. (2023), LLMs exhibit advanced capabilities in understanding and addressing complex problems through natural language processing (NLP), and their potential applications in domains requiring specialized expertise—such as finance, healthcare, and law—are increasingly recognized. A growing body of research has focused on systematically evaluating the judgment capabilities of LLMs and assessing their relevance in real-world decision-making contexts.

Recent studies aimed to quantitatively evaluate not only the performance of LLMs but also their practical applicability and inherent limitations (Chang et al., 2024; Brown, 2020). Benchmarks play a central role in these assessments by providing a consistent basis for comparing model performance across tasks and by offering objective metrics for gauging practical utility. Notable examples include benchmarks for general language understanding, commonsense reasoning, and academic knowledge, each of which has contributed to assessing the broad linguistic and reasoning capabilities of LLMs.

In finance, LLMs have been applied to a wide range of tasks, including investment strategy formulation (Li et al., 2025), risk assessment (Golec and AlabdulJalil, 2025), news analytics (Dolphin et al., 2024), and financial fraud detection (Kadam, 2024), with potential applications continuing to expand. However, most existing financial benchmarks remain centered on NLP-oriented tasks such as document summarization (Hamad et al., 2024), information extraction, and question answering (Choi et al., 2025). While these benchmarks are useful for evaluating textual comprehension and retrieval, they provide limited insight into the quantitative reasoning and structured decision-making capabilities that are essential for real-world financial management.

To address this gap, we propose a financial benchmark framework specifically designed to evaluate the capacity of LLMs to make rational and consistent judgments in realistic investment decision-making settings. In contrast to prior financial benchmarks, which are primarily language-oriented, the proposed framework is based on mathematically defined portfolio optimization problems grounded in portfolio theory. This design allows for a direct assessment of the quantitative reasoning and investment decision-making capabilities of LLMs, thereby moving beyond text generation to evaluate their potential as financial advisory and decision-support systems.

In this study, we construct a benchmark dataset comprising a diverse set of portfolio optimization problems automatically generated from portfolio theory. Each problem is presented as a multiple-choice question with four portfolio options: one representing the theoretically optimal allocation (correct answer) derived from the specified objective function and constraints, and three carefully designed distractors (alternative choices). To our knowledge, this is the first benchmark to systematically evaluate the financial decision-making capabilities of LLMs using mathematically defined portfolio optimization tasks. The framework contributes not only to the academic literature on LLM evaluation but also provides a practical foundation for assessing the applicability of LLMs in financial advisory and asset management contexts. Furthermore, we create a benchmark with a total of 9,500 questions and perform an evaluation of the three widely utilized LLMs: GPT-4, Gemini 1.5 Pro, and Llama 3.1-70B.

## 2. Related work

### 2.1 LLM evaluation

General-purpose benchmarks are primarily designed to assess the comprehensive language proficiency of LLMs. Notable examples include the Massive Multitask Language Understanding benchmark (MMLU; Hendrycks et al., 2020), HellaSwag (Zellers et al., 2019), and the AI2 Reasoning Challenge (ARC; Clark et al., 2018). MMLU comprises more than 15,000 multiple-choice questions across 57 subfields (including mathematics, physics, economics, medicine, and law) and evaluates multi-step cognitive processes such as contextual comprehension, information retrieval, conceptual reasoning, and logical decision-making. HellaSwag, by contrast, focuses on commonsense reasoning of language models. These general-purpose benchmarks have proven effective in evaluating broad language understanding and logical inference capabilities, serving as essential tools for model evaluation and performance comparison.

More recently, benchmark datasets have been developed to evaluate the financial capabilities of LLMs. For example, FinQA (Chen et al., 2021) is constructed from financial statements and requires models to answer natural language questions through reasoning and numerical computation. Rather than relying on simple information retrieval, the tasks involve calculations and numerical reasoning steps to test whether models can apply mathematical logic to financial documents. ConvFinQA (Chen et al., 2022) extends this framework to conversational settings, assessing whether LLMs can maintain contextual understanding and reasoning during conversational interactions, which are capabilities critical for applications such as robo-advisors and financial assistant chatbots. FinBEN (Xie et al., 2024) is a large-scale benchmark built from unstructured financial texts and spans 24 financial tasks ranging from information extraction, question answering, textual analysis, and risk management.

Even though these domain-specific benchmarks have advanced the evaluation of LLMs in finance, most remain concentrated on language-processing tasks. Few efforts address structured, quantitative decision-making problems (such as portfolio construction and risk-return trade-off analysis) where mathematical optimization is fundamental. As a result, existing benchmarks are limited in assessing whether LLMs possess the numerical reasoning skills and optimization-based decision-making capabilities necessary for real-world financial applications, including asset allocation and investment strategy design.

To address this gap, this study presents a financial benchmark framework based on portfolio theory. The framework generates optimization-based decision-making problems under various objectives, constraints, and market scenarios, allowing for systematic evaluation of LLMs' capacity for theory-driven portfolio decisions. This approach extends current evaluation models beyond language comprehension to quantitative decision-making, which is an essential validation step for checking the reliability of LLMs for financial applications such as portfolio management and personalized investment advisory services.

### 2.2 Portfolio theory

Portfolio theory (Markowitz, 1952), a cornerstone of modern finance, provides a mathematical framework for optimal asset allocation by jointly considering return and risk. By incorporating asset returns, volatilities, and correlations, investors can construct portfolios aligned with specific objectives such as maximizing return or minimizing risk while exploiting diversification benefits. These objectives are formulated as optimization problems with well-defined objective functions and constraints, guaranteeing the existence of optimal solutions (basic portfolio models are formulated as a convex optimization problem). This makes portfolio theory particularly well-suited for evaluating the decision-making capabilities of LLMs in investment contexts.

The mean-variance model, also referred to as mean-variance portfolio optimization, can be formulated as

$$\text{minimize}_w \quad w^{\mathrm{T}}\Sigma w \quad \text{subject to} \quad w^{\mathrm{T}}\mu \geq \mu_0, \;\; w^{\mathrm{T}}\mathbf{1} = 1, \;\; w_i \geq 0$$

where $w \in \mathbb{R}^n$ denotes the vector of portfolio weights for *n* assets, $\mu \in \mathbb{R}^n$ represents the expected return vector, $\Sigma \in \mathbb{R}^{n\times n}$ is the covariance matrix of asset returns, $\mu_0 \in \mathbb{R}$ denotes the minimum required return level, and $\mathbf{1}$ denotes the vector of ones. The objective of this optimization problem is to determine a portfolio that minimizes risk (measured as variance) while satisfying a predefined minimum expected return threshold $\mu_0$ when short-selling is restricted. When applying the portfolio model, alternative objectives such as maximizing expected portfolio return, maximizing Sharpe ratio, and minimizing conditional value-at-risk are widely used (Kim et al., 2021; Kim et al., 2024).

# 3. Methodology

## 3.1 Benchmark dataset framework

The benchmark framework proposed in this study is designed to evaluate the ability of LLMs to make optimal investment decisions under conditions that closely resemble real-world investment environments. Figure 1 illustrates the overall architecture and workflow of the proposed benchmark.

As shown in Figure 1, the user specifies the input parameters for generating benchmark problems. These include the list of investment assets (asset list), the investment objective (objective), and the investment period (start and end dates). These parameters define portfolio decision problems. For example, a user may select the objective to minimizing portfolio volatility, the list of assets to include as BND, VTI, and GSG, and set the investment period as 2020-01-01 to 2024-12-31.

Based on the specified parameters, the benchmark framework formulates and solves a portfolio optimization problem. Based on the principles of mean-variance portfolio theory, the optimization produces a mathematically optimal allocation that satisfies the given constraints (e.g., the sum of asset weights equals one, maximum allocation limits) while achieving the defined investment objective. This optimal portfolio serves as the correct answer for the benchmark question.

Since each multiple-choice question requires one correct answer and several incorrect choices (i.e., distractors), the next step is to generate incorrect distractors quantitatively by measuring their distance from the optimal solution. For example, the Euclidean distance between the optimal portfolio weights and an alternative weight vector provides a measure of their divergence. Alternative weights with larger Euclidean distances create relatively easy questions, as the incorrect options are more notably different from the optimal solution.

The framework presented in Figure 1 generates benchmark questionnaires across diverse investment settings, including variations in asset list, objective, and investment horizon. Its primary strength lies in the ability to dynamically construct evaluation problems, due to the mathematical expressiveness and flexibility of portfolio theory. This design supports the generation of problems that differ in complexity, market scenario, and investment goal, thereby allowing scalable and fine-grained evaluation of LLM decision-making performance.

In our implementation, we consider four investment objectives, five investment constraints, and four methods for generating alternative choices, as shown in Figure 2.[1] The resulting dataset comprises 9,500 questions spanning five optimization objectives (minimizing volatility, maximizing return, maximizing the Sharpe ratio, minimizing maximum drawdown (MDD), and minimizing conditional value-at-risk (CVaR)) under varying constraint conditions (no constraints, lower and upper bounds on asset weights, and fixed asset composition requirements). Distractor generation is implemented through methods such as distance-based, threshold-based, and quantile-based approaches, which are later discussed in detail.

Figure 3 illustrates an example question from the benchmark dataset. The prompt specifies the investment objective, the list of assets, the investment horizon, and instructs the model to select the portfolio that best satisfies the stated objective. This prompt requires the LLM to go beyond textual comprehension and engage in quantitative optimization reasoning. Each benchmark question consists of three elements: a natural language prompt, the optimal portfolio (correct answer), and three alternative portfolios (distractors). By combining these components, the framework generates a flexible and scalable benchmark dataset. While we created a benchmark with 9,500 questions, users can create as many questions as needed by varying asset sets, investment objectives, and investment horizons, thereby enabling multidimensional evaluation of LLM decision-making performance.

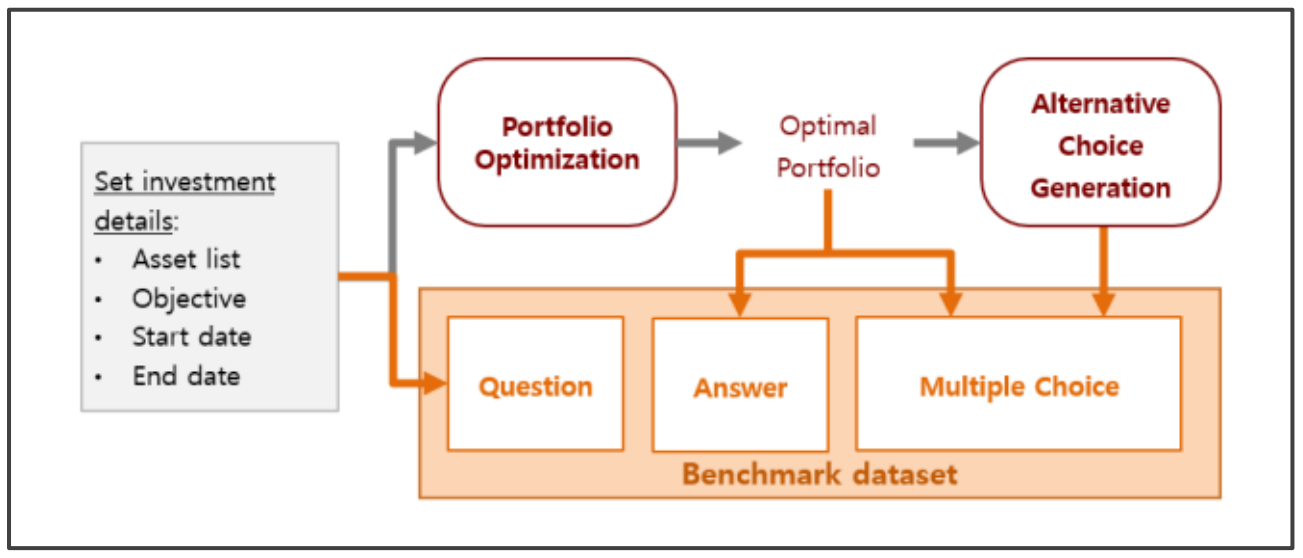

**Figure 1: Benchmark construction framework utilizing portfolio optimization**

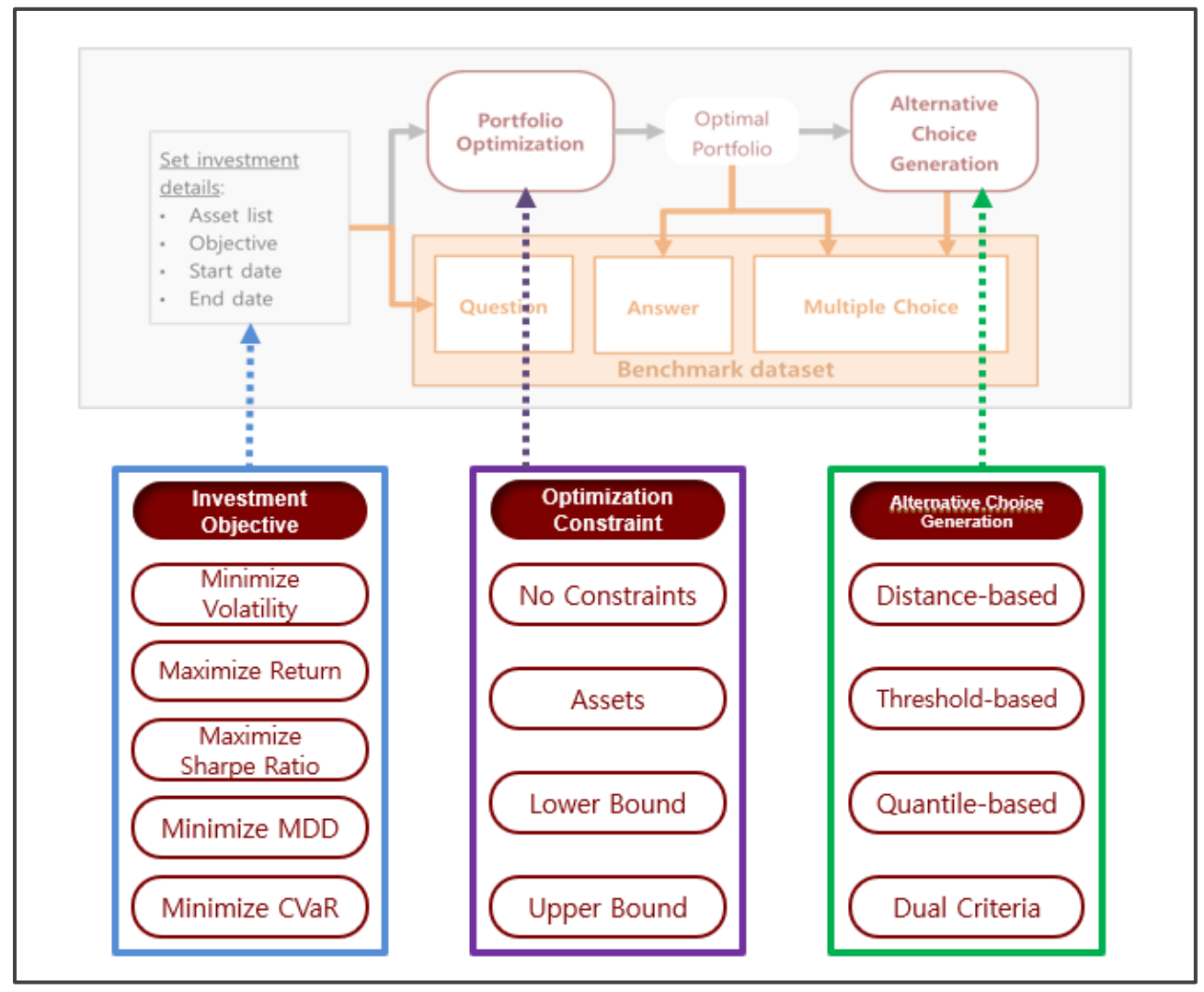

**Figure 2: Implementation of benchmark dataset for portfolio decisions**

[1] The code is available at https://github.com/noahardyx/PortBench.

We emphasize that the framework ensures each question has a mathematically unambiguous correct answer. This design eliminates the subjectivity often present in natural language-based benchmarks and allows for precise measurement of model performance. It is also important to note that the benchmark does not include highly complex allocation problems requiring optimization solvers. We fully acknowledge that an optimization agent could obtain the optimal solution if given access to an optimization solver tool. In contrast, the proposed benchmark is intended to evaluate the capabilities of foundation LLMs with respect to the basic principles of portfolio theory, including diversification and risk management.

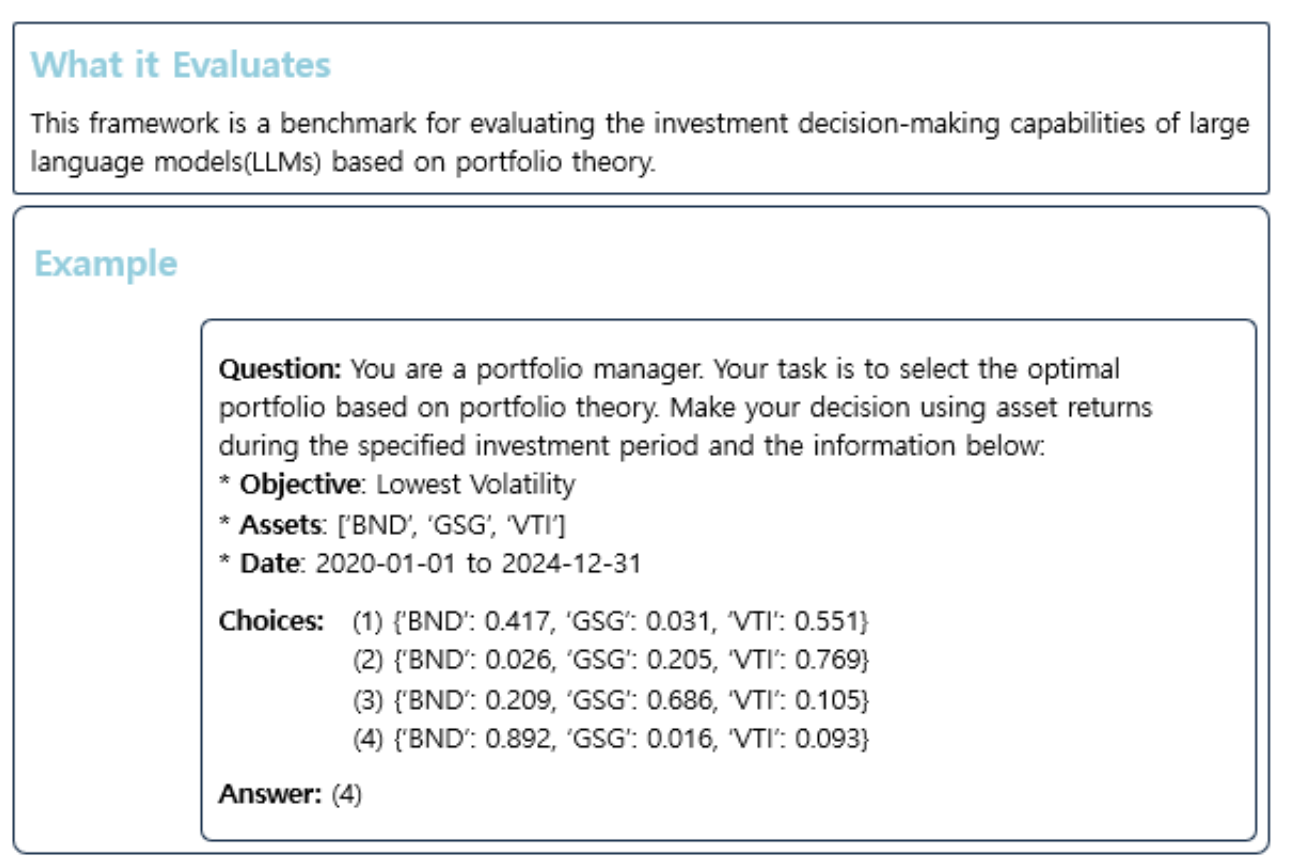

**Figure 3: Example benchmark question for portfolio decisions**

## 3.2 Distractor generation

By applying several methods for generating incorrect distractors, the benchmark incorporates questions with varying levels of difficulty, enabling more accurate evaluation of an LLM's capabilities. In our implementation, each distractor is systematically constructed based on either its structural similarity to, or performance deviation from, the optimal portfolio. Quantitative criteria and constraint conditions are applied to ensure controlled variation in difficulty while avoiding trivial or misleading alternatives. The process of generating alternative choices is organized into four distinct approaches, described below.

The first method applies a distance-based approach, where alternative portfolios are selected according to their Euclidean distance from the optimal portfolio. This method maintains a controlled degree of deviation from the correct solution, avoiding distractors that are either too similar or excessively dissimilar. Such control allows for effective adjustment of question difficulty. Candidate portfolios are retained only if their distances fall within a pre-specified range: $\theta_{min} \leq d_i \leq \theta_{max}$ where $\theta_{min}$ and $\theta_{max}$ denote the lower and upper bounds of allowable distance, respectively. In our experiment, we set distance ranges to [0, 0.25], (0.25, 0.50], (0.50, 0.75], and (0.75, 1].

The second method adopts a threshold-based approach, where alternative portfolios are selected according to their deviation from the optimal portfolio in terms of the objective function value (e.g., variance or expected return). This method evaluates not only the model's ability to recognize structural differences in asset composition but also its capacity to make performance-based judgments. A candidate portfolio is designated as a distractor if the relative difference between its performance metric and that of the optimal portfolio falls within a predefined acceptable range $[\delta_{min}, \delta_{max}]$. For constructing our benchmark, we set the ranges to [0, 0.25], (0.25, 0.50], and (0.50, 0.75].

The final method, referred to as the dual-criteria approach, generates distractors by jointly considering compositional deviation (structural similarity) and performance deviation (investment outcome) relative to the optimal portfolio. The objective is to create distractors that differ from the correct solution along both dimensions, thereby testing whether an LLM can interpret the multidimensional characteristics of portfolios. In this method, a candidate portfolio is retained as a distractor only if its Euclidean distance from the optimal portfolio lies within a specified range $[\theta_{min}, \theta_{max}]$ and its relative performance difference lies within a predefined range $[\delta_{min}, \delta_{max}]$. In addition, the weights of the alternative portfolios are constrained such that the distance between any two distractor weight vectors lies within a bounded range (e.g., 0.4 to 0.6). This requirement avoids redundancy among alternatives and ensures that each distractor exhibits distinct structural characteristics.

Furthermore, the benchmark contains questions that are generated from portfolio optimization when constraints such as lower or upper bounds (i.e., minimum or maximum allocation weights) are enforced. Lower bounds of 0%, 10%, and 20%, and upper bounds of 90%, 80%, 70%, and 60% are used in our experiment. Constraints that force portfolios to include a certain number of assets (referred to as asset constraints) are also implemented. Cases with the number of assets set to 3, 5, and 7 are used in our benchmark construction. These constraints further add variability to the dataset for evaluating LLMs.

## 4. Empirical experiment

We next present our empirical findings by evaluating investment decisions of several popular LLMs with our benchmark dataset consisting of 9,500 questions. As highlighted earlier, the benchmark was constructed using five primary investment objectives (minimizing variance, maximizing return, maximizing the Sharpe ratio, minimizing MDD, and minimizing CVaR), a variety of alternative choice generation methods (distance-based, threshold-based, quantile-based, and dual criteria) as well as several constraint settings (including no constraints, lower/upper bounds on weights, and asset cardinality constraints). In particular, we analyze three representative LLMs that are widely used and proven to perform well in various reasoning tasks: GPT-4o, Gemini 1.5 Pro, and Llama 3.1-70B. For simplicity, we refer to the three models as GPT, Gemini, and Llama, respectively. We discuss the main results and present further details in the appendix.

## 4.1 Results for investment objectives

Figure 4 compares the accuracy of LLMs across the five investment objectives. The results show that GPT attains the highest accuracy for risk-based objectives such as minimizing variance and MDD. This outcome suggests that GPT demonstrates a solid structural understanding of risk-related quantitative concepts and is capable of interpreting and applying mathematically defined objectives. In contrast, Gemini performs relatively well on return-based objectives but shows lower accuracy than GPT for other objectives. Llama recorded the lowest overall accuracy, with particularly lower accuracy compared to the other two LLMs for maximizing return and minimizing CVaR.

Overall, these findings suggest that while current LLMs can exhibit reasonably rational decision-making for single-metric objectives, they face difficulty when tasked with complex, multi-criteria optimization problems such as Sharpe ratio, which measures risk-adjusted return that must account for mean, volatility, and correlation of asset returns.

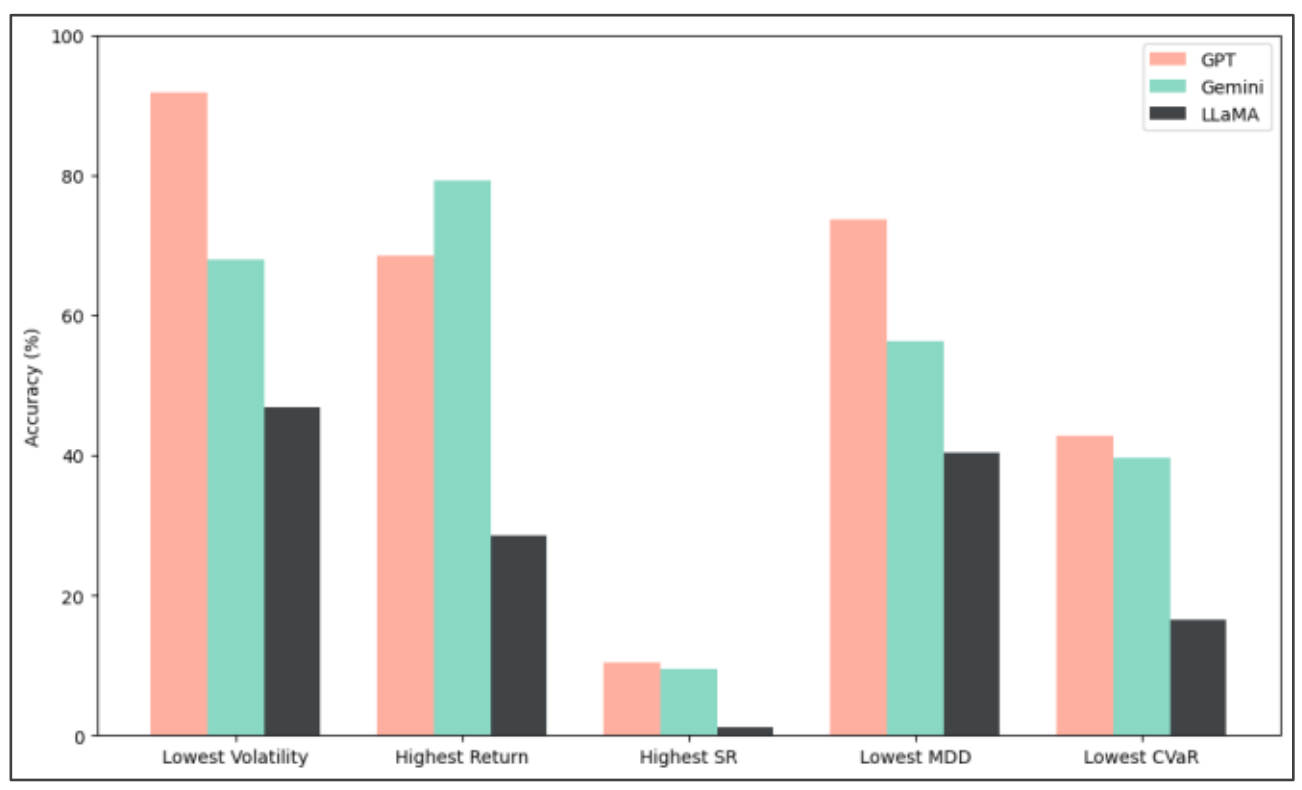

**Figure 4: Accuracy by investment objective**

## 4.2 Results for constraint type

Figure 5 presents LLMs' performance across different constraint types within each investment objective. GPT demonstrates relatively stable accuracy across all constraint settings except for Sharpe ratio maximization, with particularly strong results for minimizing variance and MDD regardless of whether constraints are imposed. Also, GPT shows high accuracy for maximizing return when no or basic constraints are imposed. This consistency indicates that GPT has a robust capacity to compare structural differences among portfolio choices, allowing it to adapt effectively to diverse decision environments. Gemini demonstrates relatively strong accuracy and especially the highest performance for return maximization. Higher accuracy is shown

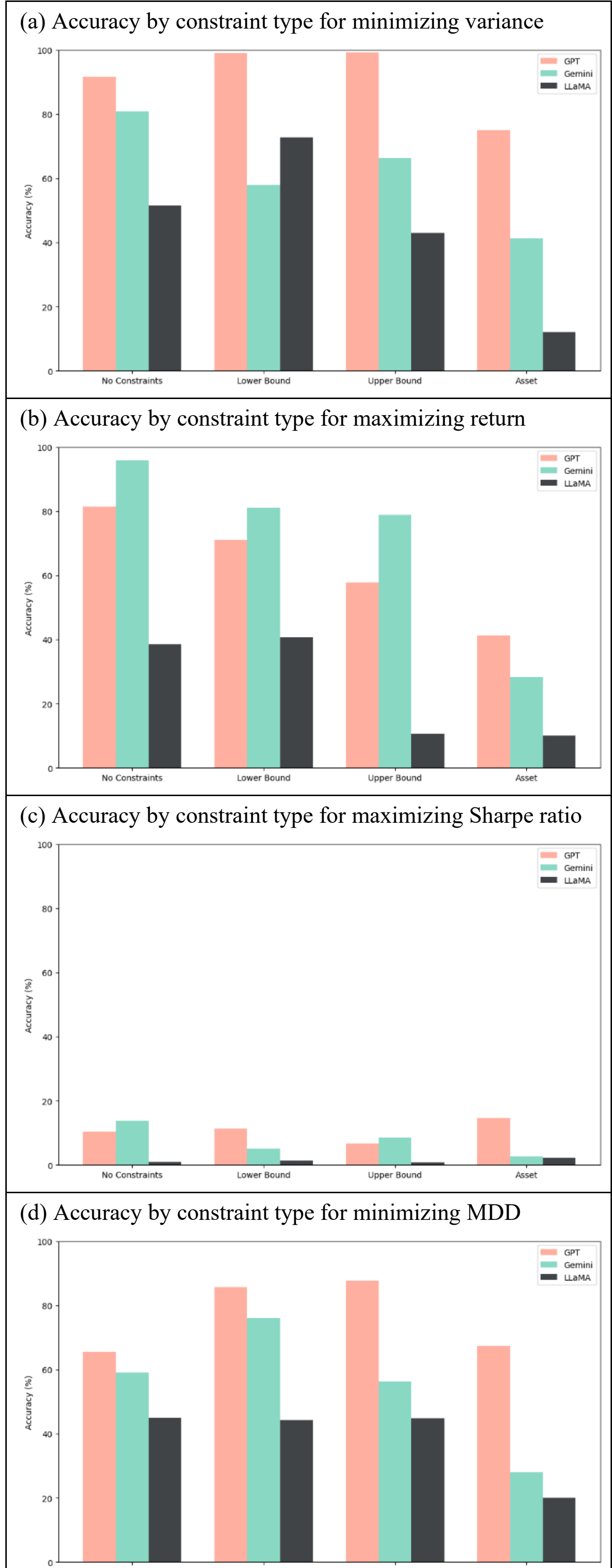

(a) Accuracy by constraint type for minimizing variance

(b) Accuracy by constraint type for maximizing return

(c) Accuracy by constraint type for maximizing Sharpe ratio

(d) Accuracy by constraint type for minimizing MDD

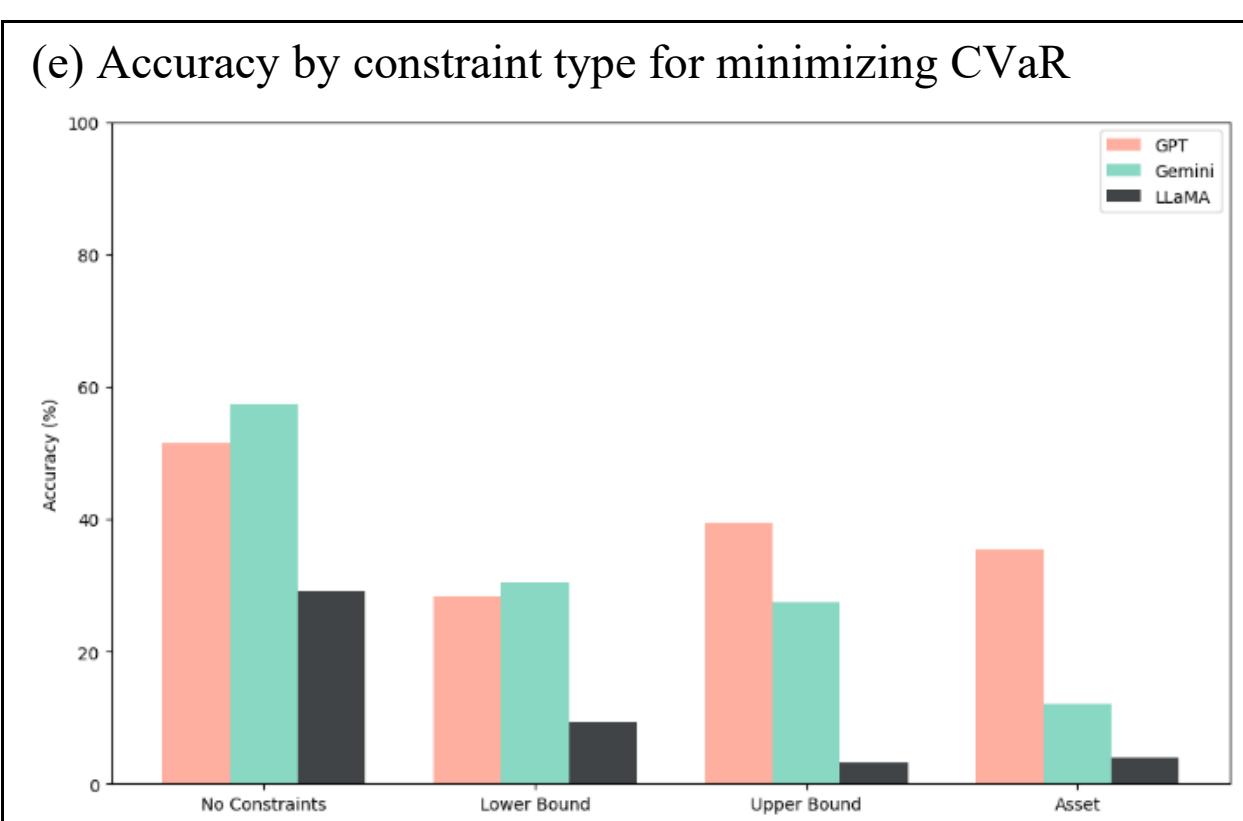

**Figure 5: Accuracy of LLMs across constraint types**

when constraints were absent or simple, but its performance declines under more complex constraints. The decline was particularly pronounced in cases involving larger asset sets and higher similarity among alternatives, suggesting that Gemini relies more on direct return comparisons than on fully interpreting structural portfolio constraints. Llama exhibits the lowest overall performance across all constraint types.

For Sharpe ratio maximization, which requires assessing the trade-off between return and risk, all three models recorded accuracy below 10% across all constraint types, showing the difficulty LLMs face when confronted with complex objective functions. More broadly, constraints had a substantial effect on portfolio selection accuracy, with performance gaps between models widening significantly under more restrictive conditions.

These results highlight that, in practical financial applications, the design of problem structures and constraints can directly influence LLM performance and must therefore be carefully considered when developing AI-driven financial decision-making systems.

## 4.3 Further discussion

The experimental results show that LLMs display distinct decision-making characteristics depending on the investment objective and the structural complexity of the problem, extending beyond simple memorization or statistical reproduction of financial knowledge. GPT consistently achieved the highest accuracy in risk-based objectives such as minimizing variance and MDD. Moreover, it maintained stable performance under varying investment constraints, indicating a relatively strong understanding of risk-related mathematical concepts and an enhanced ability to interpret and apply mathematically defined objective functions in portfolio selection tasks. In contrast, Gemini achieved relatively strong accuracy in return-based objective functions but experienced performance drops as problem complexity increased or when constraints were introduced. In particular, when distractors exhibited high similarity or when constraints were present, Gemini frequently favored portfolios with higher returns over theoretically correct solutions, leading to incorrect selections. Llama recorded the lowest overall performance, with particularly sharp declines in cases involving stronger constraints.

The key findings of this study can be summarized as follows. From a performance standpoint, GPT achieved the highest accuracy in risk-based objectives such as minimizing variance and MDD, while maintaining stable results across varying constraint conditions. This suggests that GPT has internalized structural patterns of mathematical optimization and is capable of making rational decisions consistent with clearly defined objective functions. By contrast, Gemini performed well in return-based optimization problems but showed steep performance declines in other problems, underscoring its limitations in quantitatively interpreting and applying structural constraints. Llama recorded the lowest overall performance, struggling most in tasks that combined complex objectives with restrictive constraints.

The effects of problem complexity and constraints were also evident. While all three models performed relatively well in simple, unconstrained problems, accuracy declined sharply for Gemini and Llama as constraints—such as upper and lower bounds—became more restrictive. Greater similarity among alternative choices further reduced the models' ability to differentiate differences, increasing the likelihood of incorrect portfolio selection. Performance was especially poor across all models in complex objectives such as the Sharpe ratio and CVaR, which require integration of multidimensional performance factors and quantitative interpretation of the return–risk trade-off. These findings reveal the limitations of current LLMs in mathematical reasoning when confronted with multi-objective optimization problems.

Therefore, from a practical standpoint, GPT shows strong potential as a decision-support tool for risk-focused investment strategies. However, in real-world financial settings that combine composite objectives with various constraints, none of the models demonstrate sufficient reliability for autonomous decision-making without expert validation and complementary oversight. Future advancements in LLMs should therefore aim to strengthen their capacity to perform multi-objective optimization reasoning and integrate numerical analysis with linguistic interpretation.

## 5. Conclusion

This study proposes a finance-specific benchmark framework grounded in portfolio theory to evaluate the investment decision-making capabilities of LLMs. The framework systematically generates problems based on portfolio optimization that can verify whether an LLM can select the optimal portfolio under various investment objectives and constraint conditions.

The contributions of this study are threefold. First, it introduces a benchmark framework focused on portfolio optimization problems with mathematically explicit solutions, allowing for systematic evaluation of LLMs' financial decision-making capabilities. Second, it achieves scalability and reproducibility by automatically generating problems of varying difficulty through combinations of investment objectives, asset list, investment constraints, and investment periods. Third, it provides a clear identification of each model's strengths and limitations, offering guidance for the design of LLM-based financial decision services.

Future research could extend the current multiple-choice evaluation to incorporate open-ended responses, enabling deeper analysis of LLMs' reasoning processes and conceptual understanding of portfolio theory. Moreover, integrating real market data and dynamic investment scenarios will allow for assessment of adaptability and consistency in evolving environments, offering a more realistic evaluation of the potential for LLMs as practical decision-support tools for financial management.

## ACKNOWLEDGMENTS

This work was supported by the National Research Foundation of Korea(NRF) grant funded by the Korea government(MSIT) (No. RS-2025-02216640).

## Appendix

**Table A1: Accuracy by investment objective**

| Investment objective | GPT (%) | Gemini (%) | Llama (%) |
|---|---|---|---|
| Min variance | 91.73 | 67.89 | 46.79 |
| Max return | 68.42 | 79.26 | 28.47 |
| Max Sharpe ratio | 10.47 | 9.53 | 1.21 |
| Min MDD | 73.68 | 56.21 | 40.42 |
| Min CVaR | 42.74 | 39.67 | 16.58 |

**Table A2: Accuracy by investment objective and constraints**

| | | GPT (%) | Gemini (%) | Llama (%) |
|---|---|---|---|---|
| Min variance | No Constraint | 91.56 | 80.78 | 51.44 |
| | Lower Bound | 99 | 58 | 72.67 |
| | Upper Bound | 99.25 | 66.25 | 43 |
| | Asset | 75 | 41.33 | 12 |
| Max return | No Constraint | 81.33 | 95.89 | 38.56 |
| | Lower Bound | 71 | 81 | 40.67 |
| | Upper Bound | 57.75 | 78.75 | 10.5 |
| | Asset | 41.33 | 28.33 | 10 |
| Max Sharpe ratio | No Constraint | 10.44 | 13.78 | 1 |
| | Lower Bound | 11.33 | 5 | 1.33 |
| | Upper Bound | 6.75 | 8.5 | 0.75 |
| | Asset | 14.67 | 2.67 | 2.33 |
| Min MDD | No Constraint | 65.56 | 59 | 45 |
| | Lower Bound | 85.67 | 76 | 44.33 |
| | Upper Bound | 87.75 | 56.25 | 44.75 |
| | Asset | 67,33 | 28 | 20 |
| Min CVaR | No Constraint | 51.44 | 57.44 | 29.11 |
| | Lower Bound | 28.33 | 30.33 | 9.33 |
| | Upper Bound | 39.5 | 27.5 | 3.25 |
| | Asset | 35.33 | 12 | 4 |

**Table A3: Accuracy by distance-based distractors**

| $0 \leq d \leq 0.25$ | | GPT (%) | Gemini (%) | Llama (%) |
|---|---|---|---|---|
| Distance-based | Min var | 74 | 53 | 20 |
| | Max return | 80 | 99 | 49 |
| | Max SSR | 15 | 28 | 3 |
| | Min MDD | 41 | 22 | 21 |
| | Min CVaR | 60 | 27 | 33 |
| $0.25 < d \leq 0.5$ | | GPT (%) | Gemini (%) | Llama (%) |
| Distance-based | Min var | 96 | 86 | 39 |
| | Max return | 76 | 100 | 25 |
| | Max SSR | 3 | 10 | 0 |
| | Min MDD | 64 | 68 | 48 |
| | Min CVaR | 49 | 66 | 26 |
| $0.5 < d \leq 0.75$ | | GPT (%) | Gemini (%) | Llama (%) |
| Distance-based | Min var | 96 | 98 | 71 |
| | Max return | 85 | 100 | 17 |
| | Max SSR | 6 | 6 | 0 |
| | Min MDD | 79 | 86 | 57 |
| | Min CVaR | 27 | 70 | 17 |
| $0.75 < d \leq 1$ | | GPT (%) | Gemini (%) | Llama (%) |
| Distance-based | Min var | 100 | 98 | 97 |
| | Max return | 83 | 94 | 50 |
| | Max SSR | 15 | 5 | 0 |
| | Min MDD | 89 | 90 | 80 |
| | Min CVaR | 37 | 78 | 42 |

**Table A4: Accuracy by threshold-based distractors**

| $0 \leq \delta \leq 0.25$ | | GPT (%) | Gemini (%) | Llama (%) |
|---|---|---|---|---|
| Threshold-based | Min var | 63 | 45 | 10 |
| | Max return | 82 | 98 | 37 |
| | Max SSR | 9 | 26 | 1 |
| | Min MDD | 31 | 29 | 16 |
| | Min CVaR | 55 | 24 | 21 |
| $0.25 < \delta \leq 0.5$ | | GPT (%) | Gemini (%) | Llama (%) |
| Threshold-based | Min var | 96 | 77 | 32 |
| | Max return | 81 | 96 | 25 |
| | Max SSR | 11 | 15 | 1 |
| | Min MDD | 35 | 35 | 25 |
| | Min CVaR | 63 | 52 | 44 |
| $0.5 < \delta \leq 0.75$ | | GPT (%) | Gemini (%) | Llama (%) |
| Threshold-based | Min var | 99 | 79 | 37 |
| | Max return | 86 | 93 | 45 |
| | Max SSR | 18 | 5 | 1 |
| | Min MDD | 69 | 60 | 35 |
| | Min CVaR | 49 | 62 | 24 |

**Table A5: Accuracy by quantile-based distractors**

| | | GPT (%) | Gemini (%) | Llama (%) |
|---|---|---|---|---|
| Quantile-based | Min var | 100 | 99 | 83 |
| | Max return | 84 | 87 | 48 |
| | Max SSR | 7 | 11 | 3 |
| | Min MDD | 90 | 74 | 55 |
| | Min CVaR | 50 | 77 | 23 |

**Table A6: Accuracy by lower-bound constraint**

| ≥ 0% | | GPT (%) | Gemini (%) | Llama (%) |
|---|---|---|---|---|
| Lower Bound | Min var | 100 | 95 | 82 |
| | Max return | 83 | 92 | 48 |
| | Max SSR | 2 | 14 | 0 |
| | Min MDD | 90 | 77 | 57 |
| | Min CVaR | 48 | 88 | 27 |
| ≥ 10% | | GPT (%) | Gemini (%) | Llama (%) |
| Lower Bound | Min var | 99 | 42 | 51 |
| | Max return | 71 | 88 | 22 |
| | Max SSR | 13 | 1 | 2 |
| | Min MDD | 87 | 76 | 54 |
| | Min CVaR | 20 | 0 | 1 |
| ≥ 20% | | GPT (%) | Gemini (%) | Llama (%) |
| Lower Bound | Min var | 98 | 37 | 85 |
| | Max return | 59 | 63 | 52 |
| | Max SSR | 19 | 0 | 2 |
| | Min MDD | 80 | 75 | 22 |
| | Min CVaR | 17 | 3 | 0 |

**Table A7: Accuracy by upper-bound constraint**

| ≤ 90% | | GPT (%) | Gemini (%) | Llama (%) |
|---|---|---|---|---|
| Upper Bound | Min var | 100 | 88 | 69 |
| | Max return | 67 | 81 | 12 |
| | Max SSR | 5 | 8 | 0 |
| | Min MDD | 92 | 86 | 46 |
| | Min CVaR | 44 | 38 | 8 |
| ≤ 80% | | GPT (%) | Gemini (%) | Llama (%) |
| Upper Bound | Min var | 98 | 72 | 55 |
| | Max return | 65 | 79 | 12 |
| | Max SSR | 7 | 8 | 0 |
| | Min MDD | 89 | 68 | 55 |
| | Min CVaR | 36 | 27 | 3 |
| ≤ 70% | | GPT (%) | Gemini (%) | Llama (%) |
| Upper Bound | Min var | 100 | 57 | 33 |
| | Max return | 59 | 83 | 13 |
| | Max SSR | 5 | 8 | 0 |

| | Min MDD | 90 | 44 | 40 |
|---|---|---|---|---|
| | Min CVaR | 39 | 32 | 1 |
| ≤ 60% | | GPT (%) | Gemini (%) | Llama (%) |
| Upper Bound | Min var | 99 | 48 | 15 |
| | Max return | 40 | 72 | 5 |
| | Max SSR | 10 | 10 | 3 |
| | Min MDD | 80 | 27 | 38 |
| | Min CVaR | 39 | 13 | 1 |